\documentclass[praaplied, aps, reprint, superscriptaddress,showpacs,amsmath,amssymb, twocolumn, floatfix]{revtex4-1}

\usepackage{graphicx}% Include figure files
\usepackage{dcolumn}% Align table columns on decimal point
\usepackage{bm}% bold math
\usepackage{color}
\usepackage{ulem}
\usepackage[utf8]{inputenc}
\usepackage{siunitx}
\usepackage{hyperref}% add hypertext capabilities
\definecolor{KommentarPhilip}{gray}{.5}

\begin{document}

%\preprint{PRL}

\title{Dia- and adiabatic dynamics in a phononic network}% Force line breaks with \\

\author{Daniel Schwienbacher}
\email[]{daniel.schwienbacher@wmi.badw.de}
\thanks{These two authors contributed equally}
\affiliation{Walther-Meißner-Institut, Bayerische Akademie der Wissenschaften, Walther-Meißner-Str. 8, 85748 Garching, Germany}
\affiliation{Physik Department, Technische Universität München, James-Franck-Str. 1, 85748 Garching, Germany}
\affiliation{Munich Center for Quantum Science and Technology (MCQST), Schellingstr. 4, 80799 München, Germany}
\author{Thomas Luschmann}
\thanks{These two authors contributed equally}
\affiliation{Walther-Meißner-Institut, Bayerische Akademie der Wissenschaften, Walther-Meißner-Str. 8, 85748 Garching, Germany}
\affiliation{Physik Department, Technische Universität München, James-Franck-Str. 1, 85748 Garching, Germany}
\affiliation{Munich Center for Quantum Science and Technology (MCQST), Schellingstr. 4, 80799 München, Germany}
\author{Rudolf Gross}
\altaffiliation{}
\affiliation{Walther-Meißner-Institut, Bayerische Akademie der Wissenschaften, Walther-Meißner-Str. 8, 85748 Garching, Germany}
\affiliation{Physik Department, Technische Universität München, James-Franck-Str. 1, 85748 Garching, Germany}
\affiliation{Munich Center for Quantum Science and Technology (MCQST), Schellingstr. 4, 80799 München, Germany}
\author{Hans Huebl}
\email[]{hans.huebl@wmi.badw.de}
\altaffiliation{}
\affiliation{Walther-Meißner-Institut, Bayerische Akademie der Wissenschaften, Walther-Meißner-Str. 8, 85748 Garching, Germany}
\affiliation{Physik Department, Technische Universität München, James-Franck-Str. 1, 85748 Garching, Germany}
\affiliation{Munich Center for Quantum Science and Technology (MCQST), Schellingstr. 4, 80799 München, Germany}
\date{\today}% It is always \today, today,
\begin{abstract}
  Mechanical resonator networks are currently discussed in the context of model systems giving insight into problems of condensed matter physics including effects in topological phases. Here, we discuss networks based on three high-quality factor nanomechanical string resonators (nanostrings) made from highly tensile-stressed Si$_{3}$N$_{4}$. The strings are strongly coupled via a shared support and thus can form a fully mechanical, classical multi-level system. Moreover, the individual strings are tunable in frequency, which allows one to explore their coupling behaviour using continuous wave spectroscopy and time domain techniques. In particular, such systems allow for the experimental exploration of quantum phenomena such as Landau-Zener transition dynamics by studying their classical analogues. Here, we extend the previous work performed on two coupled strings to three coupled resonators and discuss the additional features of the inter-string dynamics, such as the classical analog of Landau-Zener transition dynamics in a three-mode system.
\end{abstract}
\maketitle
\begin{figure}
	\includegraphics[scale=1]{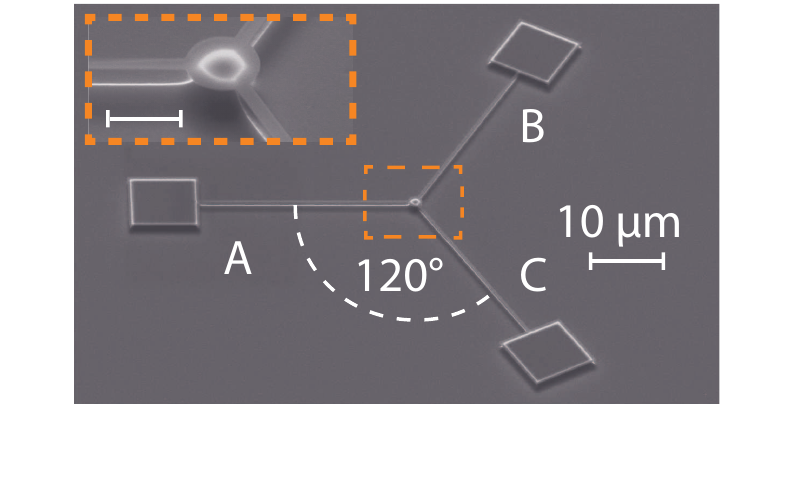}
	    \caption{
	    %a) Concept of the used in-situ tunable nano-string network, showing active energy transfer from one mode(I) to selectively one other mode (II,III), or to both modes at the same time (IV). 
	    Tilted scanning electron micrograph image of the nano-string network. Three mechanical string resonators are mechanically coupled via a disk-shaped shared support in the center of the equi-angular network. The inset shows a detailed view of the coupling region. The circular coupler (scale bar $=\SI{2}{\micro\meter}$) is partly suspended realizing the interaction between the nano-strings. 
	    %d) Schmematic of the measurement setup: The out of plane displacement of a selected  string resonator is detected using laser interferometry. Three separate RF sources control the resonance frequencies of the individual resonators in-situ. For more details see text.
	    }
    	\label{fig:setup}
\end{figure}
\begin{figure}
	\includegraphics[scale=1]{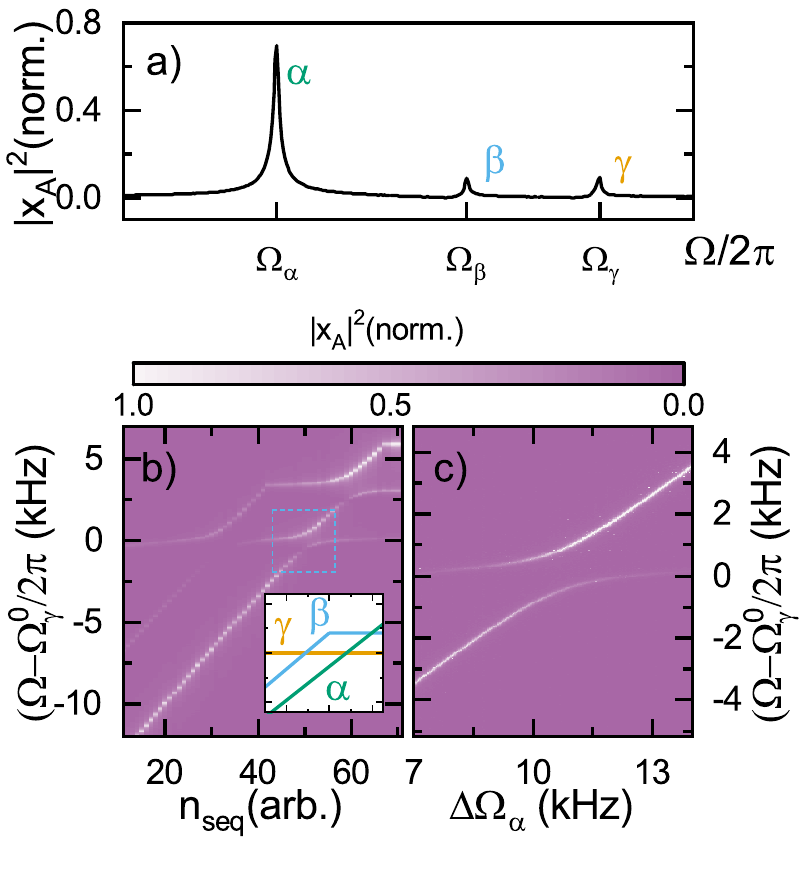}
	\caption{a) Mode distribution in the far detuned case, read out via resonator A. The modes $\alpha$, $\beta$ and $\gamma$ are distinguishable. b) Normalized mechanical response $|x_\mathrm{A}|^2$ of the string system read out via nano-string A when controlling the resonance frequencies as indicated in the inset. At points of frequency matching between two modes, avoided crossings are visible. Inset: Tuning scheme, showing the in-situ set resonance frequencies for the different modes: mode $\alpha$ (green), $\beta$ (blue), $\gamma$ (yellow) c) Thermal displacement spectrum showing the avoided crossing between the resonantly coupled modes $\alpha$ and $\gamma$. Bright areas indicate a large normalized mechanical response $|x_\mathrm{A}|^2$ of resonator A}
	\label{fig:tuning}
\end{figure}

%\section{Intro}
\label{sec:intro}
Efficient and precise control over quantum systems is one of the key goals in quantum technology and in particular quantum information processing. For the archetypal system of two strongly coupled quantum states, the Landau-Zener model provides an intuitive and analytic description of the time-dependent excitation transfer dynamics \cite{Zener1932}. Remarkably, the Landau-Zener model is not restricted to quantum states, but also successfully describes the dynamics of classical harmonic oscillators \cite{Maris1988,Shore2009,Novotny2010}. Thus, classical harmonic oscillators can be employed for modeling quantum processes under certain conditions \cite{Faust_2012,Maris1988,Shore2009,Novotny2010,Fu2016,Pernpeintner2018,Salerno2017}. However, the original Landau-Zener model is restricted to two coupled quantum states, and extending it to a larger number of states presents a challenging task \cite{Carroll1986}. Altough the three-level case can be easily defined, no explicit analytic solution exists for the general case. 
%
%Simplifications, such as treating the system as two subsequent individual two level systems \cite{Brundobler1993}, the so-called independent-crossing approximation (ICA), allow to bypass the challenge to some extent.
%
Only special cases of multi-level systems with a reduced complexity can be treated analytically \cite{Carroll1986,Ashhab2016, Militello:2019bs,Shytov:2004kr}. Here, we present an experimental approach to study and thereby simulate the complex dynamics in a controllable three-level system that exceeds the analytical description put forward in Refs.~\cite{Carroll1986,Ashhab2016,Militello:2019bs,Shytov:2004kr}. In particular, we focus on the diabatic and adiabatic state transfer between the modes. We compare the experimental results quantitatively with  numerical simulations as well as approximations such as the independent crossing  when applicable. Technically, the three-level system is realized in form of a coupled network of three silicon-nitride string resonators with high quality factors $Q$ which are mutually coupled via a shared support structure. We investigate the coupling rates and perform state transfer protocols. 
We find good agreement with numerical simulations of the full three-state Landau-Zener model, using only predetermined system parameters. We therefore explore the system as a classical simulator of the equations of motion and envision the possible future use as a simulator of a quantum state transfer processes. Besides the simulation of state transfer dynamics, coupled nano-string networks have the potential to enable a detailed understanding of the coupling rates and the controlled transfer of excitations between spatially separated nano-string resonators. This is of high relevance for sensing applications \cite{Kolkowitz2012,Rocheleau2010,Truitt2007, Biswas:2014ku}, information transport and distribution \cite{Habraken2012,Lee2004,Hoppensteadt2001}, and gives access to topologically interesting exceptional points \cite{Gao2015,Xu2017}.
%\section{Theoretical Model}
\label{sec:design}
The problem of three interacting quantum states without relaxation is in general described by the hamiltonian \cite{Militello:2019bs,Carroll1986}
\begin{equation}
   H^{(3)}= \hbar
    \begin{pmatrix}
    \Omega_{\alpha}(t) & g_{\alpha\beta} & g_{\alpha\gamma} \\
    g_{\alpha\beta} & \Omega_{\beta}(t) & g_{\beta\gamma} \\
    g_{\alpha\gamma} & g_{\beta\gamma} & \Omega_{\gamma}(t)
    \end{pmatrix}.
    \label{eq:3b3matrix}
\end{equation}
Here, the resonance frequencies of the levels or modes are $\Omega_{i}$ $(i=\alpha,\beta,\gamma)$ and the inter-mode coupling rates are given by $g_{ij}$. For the study of the excitation transfer between the modes, we introduce chirped resonance frequencies  $\Omega_{i}(t)=\Omega_{i}^0+\zeta_i \, t$, where  $\Omega_{i}^0$ is the natural mode frequency and $\zeta_i$ is the frequency ramp-rate. Later on, $\zeta_i$ will represent the control parameter to access both adiabatic and diabatic regimes of state transfer between modes \cite{Landau1932,Zener1932}. Albeit the absence of an analytic solution for the general case \cite{Carroll1986,Brundobler1993,Militello:2019bs}, analytic approximations exist for certain special scenarios such as the equal slope case $\zeta_{\alpha}\neq \zeta_{\beta}=\zeta_{\gamma}$ or the so-called bowtie case (for details see \cite{Carroll1986,Brundobler1993}).
However, both approximations assume certain inter-modal coupling rates to vanish and therefore cannot be applied to our system. 
Evidently, the systems dynamics also strongly simplify, if we can isolate the behavior of two coupled modes, i.e. by strongly detuning the remaining one (e.g. $|\Omega_\alpha-\Omega_{\beta,\gamma}|\gg \mathrm{max}(g_{\alpha\beta},g_{\alpha\gamma})$). This situation is known as the independent crossing approximation (ICA) as introduced by Brundobler et al.\,\cite{Brundobler1993}, which reduces the complexity of the three-level system to the standard Landau-Zener case \cite{Landau1932,Zener1932,Wannier1965} (cf. supplemental material \cite{SupplementaryInformation}).\\
%
%\section{Experimental details}
\label{sec:sample}
We implement the experimental realization of the three-level system in the form of three nano-mechanical string resonators sharing a disk-shaped support structure (see Fig.~\ref{fig:setup}). These high-$Q$ harmonic oscillators are freely suspended silicon nitride nano-strings labeled (A,B,C) with $Q$-factors in the $1.5 \times 10^5$ range. The fabrication starts with a single crystalline silicon wafer, which is commercially coated with a $t_\mathrm{SiN}=\SI{90}{\nm}$ thick, highly tensile-stressed, LPCVD (low pressure chemical vapor deposition) grown $\mathrm{Si_{3}}\mathrm{N_{4}}$(SiN) film. We define an aluminum hard-mask of the string network using electron beam lithography. The mask is transferred to the SiN layer using an anisotropic followed by an isotropic reactive ion etching (RIE) process. The latter releases the strings and enables a mechanical in-plane (ip) and out-of-plane (oop) motion. Last, the aluminum etching mask is removed using chemical etching. The disk-shaped central clamping pad has a diameter of $\SI{2}{\mu\meter}$ and is partially suspended (c.f. Fig.~\ref{fig:setup}~b)), enabling a mechanical coupling between the individual strings' modes with a coupling rate in the kHz range.
To investigate the resonance frequencies and linewidths of the mechanical resonators and coupling rates within the nano-string network, we use a free space optical interferometer ($\lambda = \SI{633}{\nm}$). Using a $xyz$ piezo stage, we can spatially select each of the nano-strings and optically probe their oscillation state. The displacement is encoded in the reflected light intensity, which is detected with a photo-detector. Depending on the type of measurement, we use spectral analysis or vector analysis to determine the resonance frequencies of the nano-string network. For the experiments concerning the excitation transfer, we record the data in the time domain. To avoid air-damping and maintain the high-$Q$ factors of the strings, the whole sample is operated in vacuum ($p<\SI{0.01}{\pascal}$).
We control the excitation strength and the frequency of the modes of the network with a globally applied coherent force generated by a piezo-actuator (cf. supplemental material \cite{SupplementaryInformation}). Forced excitation of a mode is achieved using a short pulse of a sinusoidal voltage signal in resonance with the specific mode. \\
For the basic characterization of the system, we focus on the spectroscopy of the modes, their controllability and their interactions. 
Figure~\ref{fig:tuning}~a) shows the excitation state in terms of the squared displacement amplitude $|x_\mathrm{A}|^2$ of nano-string A using a controlled stimulus and vector network analysis. We observe the dominant response of the mode $\alpha$, which is the natural mode of the optically probed nano-string A. In addition, we detect signatures of the $\beta$ and $\gamma$ mode (which can be associated with the basic modes of nano-string B and C) due to the intermodal coupling. Figure~\ref{fig:tuning}\,a) shows the situation of large detuning $\Delta_{ij}= \Omega_i-\Omega_j (\Delta_{ij} \gg g_{ij})$, and thus allows us to determine the basic resonance frequencies of the nanostrings A, B and C ($\Omega^{0}_\textrm{A,B,C}/2\pi = \SI{9.24385}{\mega\hertz}, \SI{9.24985}{\mega\hertz}, \SI{9.25405}{\mega\hertz}$). Complementary thermal displacement noise measurements of all resonators indicate an intrinsic linewidth of approx.~$\SI{85}{Hz}$,
corresponding to a $Q$-factor of about 150.000 at room temperature. To determine the intermodal coupling rates, we control the resonance frequencies of these three modes using a frequency tuning scheme based on the geometric non-linearity \cite{Pernpeintner2018,SupplementaryInformation}. In particular, we set the resonance frequencies of the modes according to the configuration sequence shown in the inset of Fig.~\ref{fig:tuning}~b), which is designed to realize a scenario treatable with the ICA. Fig.~\ref{fig:tuning}~b) shows the normalized squared mechanical response $|x_\mathrm{A}|^2$ of the optically investigated string A, which as seen in Fig.~\ref{fig:tuning}~a) shows signatures of the remaining two modes. As indicated in the inset, the configuration sequence controls multiple frequencies simultaneously in this measurement, which we index by the control parameter $n_\mathrm{seq}$. Given the configuration sequence depicted in the inset, we can directly determine the individual inter-modal coupling rates $g_{ij}$ by reducing the complexity to a system of two coupled harmonic oscillators (cf. supplemental material \cite{SupplementaryInformation} Eq.~S1).\\
The intermodal coupling rates can be directly extracted from the frequency evolution of the resonance signatures during the sequence: For example in Fig.~\ref{fig:tuning}b), at $n_\textrm{seq}=28$, when mode $\beta$ is nominally resonant with mode $\gamma$, we find the signature of an avoided crossing, while mode $\alpha$ is still far detuned. Equivalent scenarios are found at $n_\textrm{seq}=47$ and at $n_\textrm{seq}=59$, where avoided crossings are visible between modes $\alpha$ and $\gamma$, and $\alpha$ and $\beta$, respectively. 
Although the avoided crossing is visible in Fig.~\ref{fig:tuning}~b), we also use thermal displacement noise spectra to quantify the undisturbed coupling rates.
Figure \ref{fig:tuning}c) displays the interaction of the $\alpha$ and $\gamma$ mode around $n_\mathrm{seq}=47$ plotted as a function of the detuning $\Delta\Omega_\alpha$. Here, $\Delta\Omega_\alpha = \Omega_\alpha - \Omega_\alpha^0$ with  $\Omega_\alpha^0$ being the undisturbed frequency of mode $\alpha$. In this particular case, the modes $\alpha$ and $\gamma$ hybridize and we can identify the dressed state frequencies $\Omega_+^{\alpha,\gamma}$ and $\Omega_-^{\alpha,\gamma}$ as the respective upper and lower branches of the avoided crossing (cf. supplemental material \cite{SupplementaryInformation}). At the point of optimal mode mixing, we can extract the mode spacing as $2g_{\alpha\gamma}/2\pi = \SI{1307}{\hertz}$ (cf. supplemental material \cite{SupplementaryInformation} Eq.~SI2) setting this coupled modes well in the strong coupling regime ($g_{\alpha\gamma} \gg \Gamma_\alpha , \Gamma_\gamma$). As the $\beta$ mode is detuned from these degenerate modes by $\SI{15}{\kilo\hertz}$, we are also well within the limit of the independent crossing approximation.
All individual inter-modal coupling rates were determined accordingly and are summarized in the supplemental material \cite{SupplementaryInformation}.\\
\begin{figure}
	\includegraphics[scale=1]{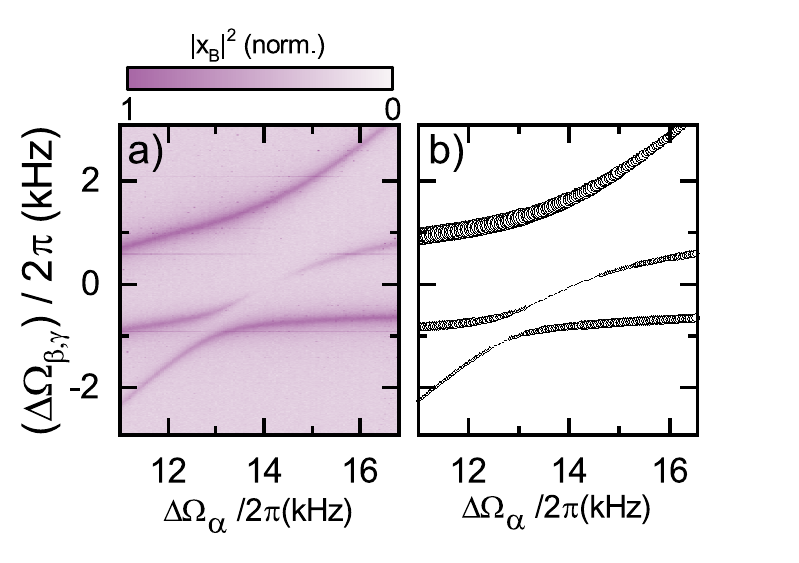}
	\caption{a) Thermal displacement spectrum of nano-string B showing the normalized mechanical response $|x_\mathrm{B}|^2$ during a tuning sequence which tunes all three modes to a degenerate frequency. The lower mode of the coupled system ($(\beta,\gamma)^-$) goes dark for a specific de-tuning combination. To allow easier comparison, the $y$-axis is centered around the initial value of the lower branch of the avoided crossing between the modes $\beta$ and $\gamma$ with $\Delta\Omega_{\beta,\gamma}=\Omega - \Omega^{\beta,\gamma}_- (\Delta \Omega_\alpha =0)$. On the $x$- axis we plot the detuning of the in-situ tuned mode $\alpha$. Here $\Delta\Omega_\alpha =\Omega_\alpha- \Omega_\alpha^0$ with $\Omega_\alpha^0$ being the initial value of the frequency of mode $\alpha$. b) Numerical simulation of the resonance frequencies from the measurement shown in a). The area of the data-points shows the amplitude of the mode at a given point in the spectrum. A bias was added to make small amplitudes, especially in the range of the dark-mode, visible.}
	\label{fig:darkmode}
\end{figure}
In the next step, we configure the frequencies of all three modes that they become degenerate. Figure~\ref{fig:darkmode}~a) shows the measured thermal displacement spectrum under these conditions. Initially, at $\Delta\Omega_\alpha=0$, the resonance frequencies are set to prepare the modes $\beta$ and $\gamma$ in a hybridized state with the normal mode frequencies $\Omega_+^{\beta,\gamma}$ and $\Omega_-^{\beta,\gamma}$. We then increase the frequency of mode $\alpha$ towards $\Omega_+^{\beta,\gamma}$ and $\Omega_-^{\beta,\gamma}$ until we reach the region of interest, where $\Omega_{\alpha}\approx\Omega_{\beta}\approx\Omega_{\gamma}$, and investigate the characteristic spectrum by optically probing the displacement of nano-string B.
At $\Delta\Omega_{\alpha} = 0$, the assumptions of the independent crossing approximation are fulfilled (for the $\beta$ and $\gamma$ mode) and an analytic description of the spectrum is possible. However, this changes as soon as $\alpha$ starts to interact with the hybridized modes ($\Omega_{\alpha}^0+\Delta\Omega_{\alpha}\lessapprox \Omega^{\beta,\gamma}_-$). This becomes apparent from the formation of an avoided crossing between mode $\alpha$ and the lower branch of the initially hybridized $\beta$ and $\gamma$ modes and the shift of the $\Omega_+^{\beta,\gamma}$ branch to higher frequencies. Notably, the mode spacing between $\Omega_\alpha$ and $\Omega_-^{\beta,\gamma}$ shows a reduced effective coupling rate of $\SI{497\pm1}{\hertz}$ at $\Delta\Omega_\alpha / 2\pi =\SI{12.9}{\kilo\hertz}$ compared to the undisturbed mode coupling rates. At around  $\Delta\Omega_\alpha/2\pi = [ \SI{11}{\kilo\hertz}]$, we find a suppression of the thermal displacement of one of the modes, i.e. a dark state. Here, the uncoupled frequencies of mode $\alpha$, $\beta$, and $\gamma$ become degenerate and the central of the three hybridized modes quenches. While analytic models cannot grasp the full richness of the spectrum, numerical simulations can be used to predict the frequency and amplitude evolution as shown in Fig.~\ref{fig:darkmode}~b). Here we have encoded the expected displacement noise amplitude in the size of the symbols. Given the independently determined coupling strengths, we find an excellent agreement between our observations and modelling.  In the numerical simulation we included a linear correction attributed to the heating induced by the piezo-actuator(cf. supplemental material \cite{SupplementaryInformation}).
While the emergence of dark states has been studied extensively in optomechanics \cite{Dong2012,Wang2012,Tian2012}, in micro-mechanical systems they have only been reported for three-level $\bigwedge$-type systems \cite{Okamoto:2016bm}. We note that we observe a dark state in a more general three level system, where all inter-modal couplings are present and comparable, thus increasing the range of observed phenomena in micro and nano-mechanical research.
%\section{state transfer}

\begin{figure}
  \includegraphics[scale=1]{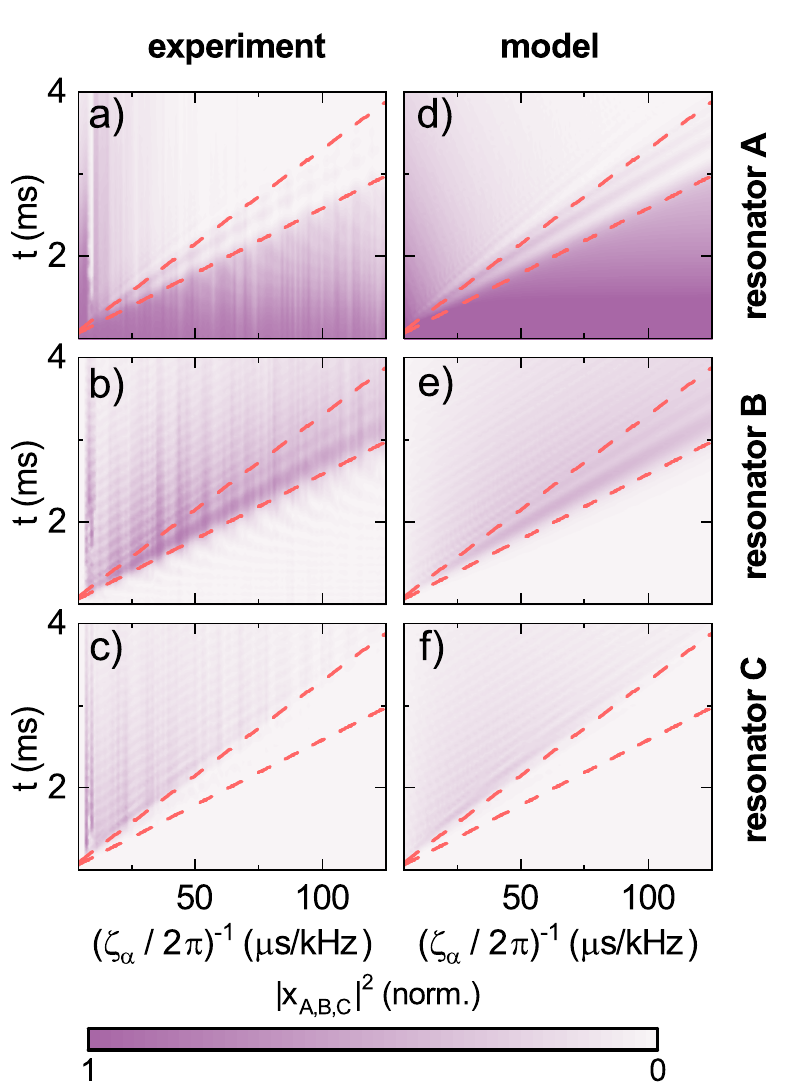}
  \caption{Excitation of the single resonators in the network over time $t$ for different tuning rates $\zeta_\alpha$ during a Landau-Zener experiment. Dark color indicates a large excitation. (a-c) show the experimental data for each string and corresponding numerical simulation results are shown in (d-f). In both cases an initial pulse is used to excite mode $\alpha$. The resonance frequency of $\alpha$ is then chirped, upwards through the resonance frequencies of the other modes with varying tuning speeds (fast to slow chirp rates). The dashed red lines are a guide to the eye to mark the times at which the resonance frequency $\Omega_\alpha$ of mode $\alpha$ matches the mode frequencies $\Omega^{\beta,\gamma}_-$ (lower) and $\Omega^{\beta,\gamma}_+$ (upper), respectively. The measurement and model are shown for all read-out resonators (A,B,C). A high mechanical response of the measured string is depicted by dark color. Each column of the graph is normalized to the initial excitation pulse.}
  \label{fig:landau-zener}
\end{figure}
Next, we turn to investigate the dynamics of this multi-mode system using experiments inspired by Landau-Zener type state transfers  \cite{Faust_2012,Fu2016,Seitner2016,Gajo2017,Pernpeintner2018}. For this, we set the frequencies of the $\beta$ and $\gamma$ mode in a static fashion. The dynamics is then probed by tuning the $\alpha$ mode
to be off-resonant with the $\beta$ and $\gamma$ mode and preparing the amplitude state of the $\alpha$ mode with a sharp excitation pulse of duration of  $t_p =\SI{400}{\micro\second}$. After a waiting period of $t_0=\SI{1}{\milli\second}$, the mode frequency of $\alpha$ is chirped towards higher frequencies with the rate $\zeta$ through the frequency range, where an interaction with the $\beta$ mode, the $\gamma$ mode or a hybridized state of $\beta$ and $\gamma$ is anticipated ($\Omega_\alpha(t)=\Omega_{\alpha}^0+\zeta_\alpha t$; for details see supplemental material \cite{SupplementaryInformation}).
To quantify the dynamics of the excitation transfer, we record the response of the displacements of nano-strings A, B, and C sequentially in the time-domain for various inverse chirp-rates ($(\zeta/2\pi)^{-1}= \SI{5}-\SI{150}{\micro\second/ \kilo\hertz}$). We distinguish two principal cases when chirping mode $\alpha$: (i) modes $\beta$ and $\gamma$ are far detuned from each other ($|\Delta_{\beta\gamma}|\gg g_{\beta,\gamma}$) representing the ICA scenario and (ii) $\beta$ and $\gamma$ are resonant and thus hybridized ($\Omega_\beta,\Omega_\gamma \rightarrow \Omega^{\beta,\gamma}_+,\Omega^{\beta,\gamma}_-$ with $\Omega^{\beta,\gamma}_+ -\Omega^{\beta,\gamma}_- \approx \SI{1.2}{kHz}$). While case (i) results in the expected (potentially sequential) Landau-Zener type state transfer depending on the adiabaticity of the process (for details see SI), the situation in case (ii) is more involved.  The data of this scenario is presented in Figure \ref{fig:landau-zener}, where the rows compare the excitation state detected on the strings A, B and C. All panels show the measured amplitude of displacement color encoded as function of the evolution time $t$ and the inverse chirp-rate $\zeta$. The dashed lines are guides to the eye indicating the time at which $\Omega_+^{\beta,\gamma}$ and $\Omega_-^{\beta,\gamma}$ become degenerate with $\alpha$. 
For small times $t$, the excitation remains on the mode $\alpha$, where it has been initially prepared, and experiences the expected exponential relaxation. Consequently, we observe no visible initial excitations on strings B and C.
For large chirp rates $(\zeta_\alpha /2\pi)^{-1}\leq \SI{7}{\micro \second / \kilo\hertz}$, representing the diabatic limit of the Landau-Zener process, the excitation remains on nano-string A (or mode $\alpha$), even after passage through the avoided crossings with both hybridized modes. In contrast, slow chirp rates $(\zeta_\alpha /2\pi)^{-1}\geq \SI{7}{\micro \second / \kilo\hertz}$ result in an adiabatic transfer to the hybridized modes, visible as the appearance of displacement signals on nano-strings B and C.
For the time interval between the passage of $\Omega_-^{\beta,\gamma}$ and $\Omega_+^{\beta,\gamma}$, we observe a slow beating in the excitation of nanostring A and B (see panels a and b)). We associate this temporal evolution with the complex interplay of the various coupling rates (cf. Fig.~\ref{fig:darkmode}). For times after the transit of both, the $\Omega_-^{\beta,\gamma}$ and $\Omega_+^{\beta,\gamma}$ mode, we find that most of the excitation in resonator A is absent and record fast beating of the excitation between resonator B and C. The latter can be understood as the pure modes of resonator B and C being hybridized and hence the excitation starts to oscillate between the physical nano-strings. This underlines the hybrid character of modes, being distributed over parts of the network. We determine this oscillation frequency  (see Ref.~\cite{SupplementaryInformation}) to $\SI{3.27}{\kilo\hertz}$, agreeing well with the experimentally set frequency separation of the coupled modes in this experiment. This finding confirms that this feature corresponds indeed to a classical Rabi oscillation. In addition, we point out that the observation of these beatings represent a clear fingerprint for the breakdown of the ICA. 
In more detail, we can compare this experimental data quantitatively to a numerical simulation of the excitation obtained via the equations of motion using the independently determined system parameters and no free fit parameters (cf. SI\cite{SupplementaryInformation}). Results are presented in panel d)-f), showing an excellent agreement of experimental data and simulation.  
%

%

%\section{summary and Outlook}
In summary, we experimentally investigated a nano-string network comprised of three nano-mechanical resonators with independently tunable resonance frequencies. This configuration allows us to test the complex problem of a fully coupled three-level system and examine configurations with and without the existence of an analytic description. We explored the formation of dark states of a fully coupled network, which are of relevance in the context of information storage, as they are considered to offer an improved robustness against noise. Other uses are for example in the field of topological energy transfer \cite{Gao2015,Xu2017,Rosenthal2018}. 
In addition, we investigated the dynamical evolution of the three interacting nano-strings and observed excitation transfer governed by the same Landau-Zener physics that is used to describe quantum state transfer processes. As such, the system represents a classical simulator of Landau-Zener dynamics in three-level systems, whose results could in principle help understand quantum state transfer processes in similarly composed quantum systems. Thus, we present a scalable path towards devices hosting multiple coupled resonators, and thus allow to simulate multiple energy levels. Finally, the investigated three-nano-string device can also be imagined as an intersection-like building block in a larger network of (nano-)mechanical devices. Therefore, we expect our findings to be relevant to the controlled transfer and routing of propagating phonons within phononic networks \cite{Habraken2012}.
%\section{Acknowledgements}
\begin{acknowledgments}
We acknowledge funding from the European Union's Horizon 2020 research and innovation program under grant agreement No 736943 and from the Deutsche Forschungsgemeinschaft (DFG, German Research Foundation) under Germany’s Excellence Strategy EXC-2111-390814868.
\end{acknowledgments}
%\bibliography{PhonSwitchBib}


%
\clearpage

\end{document}